\documentclass[sigconf] {acmart}

\AtBeginDocument{%
  }
\usepackage{ragged2e}
\usepackage{array}
\usepackage{hyperref}
\usepackage{url}
\usepackage[utf8]{inputenc}
\usepackage{enumitem}
\usepackage{orcidlink}

\copyrightyear{2024}
\acmYear{2024}
\setcopyright{rightsretained}
\acmConference[CSCW Companion '24]{Companion of the 2024 Computer-Supported Cooperative Work and Social Computing}{November 9--13, 2024}{San Jose, Costa Rica}
\acmBooktitle{Companion of the 2024 Computer-Supported Cooperative Work and Social Computing (CSCW Companion '24), November 9--13, 2024, San Jose, Costa Rica}\acmDOI{10.1145/3678884.3681894}
\acmISBN{979-8-4007-1114-5/24/11}

\makeatletter
\gdef\@copyrightpermission{
  \begin{minipage}{0.3\columnwidth}
   \href{https://creativecommons.org/licenses/by/4.0/}{\includegraphics[width=0.90\textwidth]{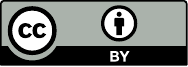}}
   
  \end{minipage}\hfill
  \begin{minipage}{0.7\columnwidth}
   \href{https://creativecommons.org/licenses/by/4.0/}{This work is licensed under a Creative Commons Attribution International 4.0 License.}
  \end{minipage}
  \vspace{5pt}
}
\makeatother

\begin{document}

\title{Enhancing Personalised Cybersecurity Guidance for Older Adults in Ireland}

\author{Ashley Sheil}
\orcid{0000-0001-7750-9495}
\affiliation{%
  \institution{Munster Technological University}
  \city{Cork}
  \country{Ireland}
  }  
\email{ashley.sheil@mtu.ie}

\author{Jacob Camilleri}
\orcid{0009-0009-4912-258X}
\affiliation{%
  \institution{Munster Technological University}
  \city{Cork}
  \country{Ireland}
}
\email{jacob.camilleri@mtu.ie}

\author{Moya Cronin}
\orcid{0009-0008-2831-186X}
\affiliation{%
  \institution{Munster Technological University}
  \city{Cork}
  \country{Ireland}
}
\email{moya.cronin@mycit.ie}

\author{Melanie Gruben}
\orcid{0009-0008-9332-5205}
\affiliation{%
  \institution{Munster Technological University}
  \city{Cork}
  \country{Ireland}
}
\email{melanie.gruben@mtu.ie}

\author{Michelle O Keefe}
\orcid{0009-0008-7130-9941}
\affiliation{%
  \institution{Munster Technological University}
  \city{Cork}
  \country{Ireland}
}
\email{michelle.okeeffe2@mtu.ie}

\author{Hazel Murray}
\authornotemark[1]
\orcid{0000-0002-5349-4011}
\affiliation{%
 \institution{Munster Technological University}
 \city{Cork}
 \country{Ireland}}
 \email{hazel.murray@mtu.ie}

\author{Sanchari Das}
\authornote{Both authors contributed equally to this research.}
\orcid{0000-0003-1299-7867}
\affiliation{%
  \institution{University of Denver}
  \city{Denver, Colorado}
  \country{USA}}
\email{sanchari.das@du.edu}

\renewcommand{\shortauthors}{Ashley Sheil et al.}

\begin{abstract}
The term \lq\lq Digital Divide\rq\rq\ emerged in the mid-1990s, highlighting the gap between those with access to emerging information technologies and those without. This gap persists for older adults even in the 21st century. To address this, our study focused on how older adults in Ireland can feel safer online. We conducted a two-phase study. In Phase I, $58$ participants used Dot Voting to identify top cyber-security priorities, including password management, privacy, and avoiding scams. This informed Phase II, where we held focus groups with $31$ participants from rural and urban communities in Ireland. Researchers provided tailored advice through presentations and leaflets, followed by open discussions. Our findings show that, despite being highly aware of cyber-scams, older adults remain very concerned about them. Participants expressed hesitation about using online password managers and two-factor authentication but valued advice on privacy and tools that can help them feel more in control online.
\end{abstract}

\ccsdesc[500]{Security and privacy~Human and societal aspects of security and privacy}
\ccsdesc[500]{Security and privacy~Social aspects of security and privacy}
\ccsdesc[500]{Human-centered computing~Accessibility}
\ccsdesc[500]{Social and professional topics~Seniors}

\keywords{Digital Divide; Older Adults; Mental Models;  Ireland; Cybersafety.}

\maketitle

\section{Introduction \& Related Work}
\label{Sec:Intro}
In Ireland, the Digital Ireland Framework~\cite{HarnessingDigital2022} and the Adult Literacy for Life Strategy~\cite{adultliteracyforlife} aim to equip the population with essential digital skills, emphasising cyber-security awareness. However, a significant digital divide persists among older adults: 65\% of those over 65 face digital exclusion, and 25\% of individuals aged 60-74 do not use the internet at all~\cite{age-action-digital-divide}. The complexity and often conflicting nature of security advice can deter engagement and hinder older adults' understanding of essential security practices~\cite{murray2023costs,hazels-thesis, murray2017evaluating, redmiles2020comprehensive}. In terms of the age demographic, the digital divide~\cite{digital-divide-origin,srinuan2011understanding} separates the generations of people who grew up with access to technology and those who did not. Barriers such as limited access to devices, internet connectivity disparities, and varying digital literacy~\cite{friemel2016digital} worsen social isolation and limit access to essential services and information~\cite{mubarak2022elderly}. COVID-19 has further exacerbated this divide~\cite{barnils2022grey,lythreatis2022digital,ramsetty2020impact,tomczyk2022digital,joshi2020substituting}. Improving older adults' digital skills is key to bridging this divide. Understanding how they engage with various technologies is crucial~\cite{zhao2023narrowing,das2019don,kebede2022digital}.

Numerous studies highlight the unique vulnerabilities of older adults to internet fraud, emphasizing the need for tailored prevention and intervention measures~\cite{judges2017role, das2020risk, shao2019older, moore2022digital, burton2022exploring, shang2022psychology, o2021can, das2019all, gopavaram2021cross}. Short-term memory and cognitive function contribute to older adults' susceptibility to cyber-attacks~\cite{ebner2020uncovering, james2014correlates, das2020non}, as well as their ability to manage passwords and maintain privacy online~\cite{zezulak2023sok}. Navigating online security advice poses challenges for users, especially older adults. Frik et al. interviewed $46$ older adults, revealing their heightened privacy and security risks due to misconceptions and usability issues with current controls~\cite{frik2019privacy}. This is evident in the literature; Murray and Malone found significant ambiguity in online security advice, with 41\% being contradictory and not reflective of best practices~\cite{murray2023costs}. A comparison study conducted by Parti showed that older victims differ from younger ones and require specific solutions~\cite{parti2021}. Kropczynski et al. emphasise tailoring digital care-giving tools to generational needs~\cite{kropczynski2021examining}. Additionally, Karagiannopoulos et al. advocate co-designing educational programs with end-users for relevance and effectiveness~\cite{karagiannopoulos2021cybercrime}. 

The research discussed in this section highlights the importance of addressing specific needs and designing educational tools to specifically bridge the gaps in older adults' knowledge, to which this study contributes. Tailored initiatives focusing on access, digital skill enhancement, and inclusivity are needed to improve older adults' quality of life in our digital world~\cite{seifert2018seniors, quan2018dividing, morris2007internet, csaki2013use, aggarwal2020impact}.

\begin{table*}[htp!]

\resizebox{0.6\textwidth}{!}{%
\begin{tabular}{@{}lr@{}}
\toprule
\multicolumn{2}{l}{\textbf{Q: Which of the following areas would you most like assistance with?}} \\ \midrule
\multicolumn{1}{l}{\textbf{Options}}                                     & \multicolumn{1}{r}{\textbf{Votes}} \\ \midrule

\textbf{A:} Online banking                                                           & 3                                  \\
\textbf{B:} Online shopping                                                          & 4                                  \\
\textbf{C:} Privacy of personal   information                                        & 8                                  \\
\textbf{D:} Password management                                                      & 8                                  \\
\textbf{E:} How to recognise a scam or   cyber-attack                                & 29                                 \\
\textbf{F:} Online communication with family and friends                           & 5                         
\\ \bottomrule
\end{tabular}%
}

\caption{Phase One; Dot vote question and options.}
\label{tab:questions-votes}
\end{table*}

To help address this issue, we have created a two-phase study to develop personalized cyber-security advice tailored specifically for older adults. In our preliminary study, more than $80$ adults participated in interviews, sharing insights into their technological experiences, cyber-security awareness, and anticipated future needs in terms of cyber-safety. Drawing from the outcomes of these interviews, several recurring themes emerged. Subsequently, the project progressed to identify the areas of cyber-safety that hold greater importance for older adults. This process involved developing a priority voting system where our participants voted on the areas that were more important to them. Our study engages older adults to determine effective tailored cyber-advice, to ensure they feel confident navigating the internet safely. The objective of this study is to bridge the digital gap for older adults in Ireland. In a two-phased approach, we aim to help address this gap by answering the following research questions:
\\
\textbf{\textit{RQ1}: Which specific cyber-security topics and practices do older adults prioritise learning about to enhance their online safety and security awareness?}\\
\textbf{\textit{RQ2}: Do older adults find such cyber-security advice and guidance helpful in improving their understanding and management of online security?} \\
\textbf{\textit{RQ3}: What additional information or resources would they be interested in learning to further enhance their cyber-security knowledge?} \\
Our contribution lies in identifying and addressing the unique cyber-safety needs of older adults, and in creating and evaluating educational materials specifically tailored to these needs. We achieve this by employing two qualitative methods: dot voting and focus groups. The results from our study can inform the development of effective cyber-education programs for older adults and guide future qualitative research in this area.

\section{Phase One: Priority Dot Voting}
\begin{center}

\end{center}
\label{dot-voting}
In order to answer \textbf{\textit{RQ1}}, we contacted several older adult groups nationwide, each with varying levels of computer proficiency. The goal was to vote on a range of different cyber-security topics highlighted in our previous interviews (mentioned in Sec.\ref{Sec:Intro}). For the voting process, we used the dot voting method~\cite{dalton2019dot}, which is a common method used in UX design~\cite{dot-voting}. A total of six groups were invited to participate in the vote, comprising of a Senior Citizens' Association, a Silver Surfers digital hub group, a Digital Skills course, two Fit Farmers' groups, and a Men's Shed group. The collective number of older adults involved was 58, with ages ranging from 59 to 86, representing a diverse range of experiences and backgrounds. The groups consisted of both male and female members.

Individual sheets were printed with our question, featuring a selection of different options. The question was presented in large font on one A4-sized page, with the options printed on additional sheets. Each participant received dot stickers to indicate their preferred option for the question presented to them. During the voting session, we instructed each group to divide into smaller units and provided them with the question and options. Participants were encouraged to deliberate on the task collectively or seek clarification as needed. The group activity created an environment to discuss the topic in a less formal manner than separate interviews, fostering a relaxed atmosphere to openly ask questions if needed. The question and its options can be viewed in Table~\ref{tab:questions-votes}. Notes were taken on the voting process and discussion, but they were not recorded. No thematic analysis was conducted in this phase of the study, as the purpose was to attain votes for phase two.

\begin{table*}[htp!]
\centering
\small
\begin{tabular}{m{5cm}m{10cm}}
\toprule
\textbf{Cyber-safety Advice} & \textbf{Recommendations} \\ \midrule
\textbf{How to avoid scams} & 

\begin{itemize}
    \item Verify the source (e.g., Bank, An Post) by calling or visiting in person.
    \item Check the legitimacy of links at www.check.cyberskills.ie\dag.
\end{itemize} \\ \midrule
\textbf{Password Management} & 
\begin{itemize}
    \item Keep passwords written down in a safe place.
    \item Never reveal your passwords.
    \item Use unique passwords with at least 12 characters for important accounts.
    \item Consider enabling two-factor authentication (2FA).
\end{itemize} \\ \midrule
\textbf{Privacy and Cookies} & 
\begin{itemize}
    \item Decline (reject all) cookies in cookie banners. If 'decline' is not available, go to settings and save preferences.
    \item Check if personal information is mandatory for forms.
    \item Avoid sharing personal details on social media.
\end{itemize} \\ \midrule
\textbf{What to Do if You are Hacked} & 
\begin{itemize}
    \item Contact your service provider (e.g., Bank, An Post).
    \item Report the crime at your local Garda station and request a PULSE number*.
    \item Provide the PULSE number to your service provider.
    \item Immediately change compromised passwords.
\end{itemize} \\ \bottomrule
\end{tabular}
\caption{Phase Two; Cyber-safety advice and recommendations presented to focus groups.
*PULSE number is a computer-generated identifier for a crime or incident, used by the Gardaí to track the case. \dag Check My Link is a national service led by Cyber Skills at Munster Technological University, in collaboration with Scam Adviser and An Garda Síochána. It aims to boost consumer confidence in the authenticity of online shopping sites and ensure they are malware-free.
}
\label{tab:cyber-advice}
\end{table*}

\subsection{Phase One: Results \& Discussion}

We found that Option E, how to recognise a cyber-attack or scam, gained the most votes totalling 29 (see Table \ref{tab:questions-votes}). The least popular choice was option A, online banking, with 3 votes. This was surprising, as online banking appeared to be of significant concern for older adults in our previous surveys. Online shopping was also valued low at 4 votes, with online privacy and password management tied at 8 votes. The number of votes received for each option can be seen in Table~\ref{tab:questions-votes}.
Aside from answering our research questions, the voting exercise sparked many discussions on the topic of cyber-safety. A prevalent worry revolved around cyber-scams. One participant inquired if there are scams tailored specifically for older individuals. While many participants demonstrated adeptness in identifying scams, they expressed apprehension about the increasing complexity of cyber-threats. One individual was perplexed by how strangers obtained access to his private landline number. Passwords were also a topic of discussion, with frustration voiced over forgetting them and the subsequent need to reset them. The groups with more technical expertise, such as the digital skills group, showed curiosity about topics like AI and concerns regarding safety and privacy.

It became evident that not only is there a technological divide between younger and older generations~\cite{mubarak2022elderly}, there is also a notable gap in knowledge within the older demographic itself. Some older adults exhibit enthusiasm and proficiency in technology, while others show disinterest and resign themselves to being excluded from the digital era. Online banking exemplifies this divide; Half of the participants reported using online banking without any issues, while the other half mentioned they never use it and had no intention of doing so in the future. These individuals believe they are incapable of learning new and complex information at their age. Consequently, these individuals encountered the most difficulty with the task and tended to vote arbitrarily. Another instance illustrating this technological disparity was revealed when a participant voiced her desire for instructional videos on camera usage and editing, underscoring her strong interest in the topic. This contrasted sharply with other group members who possessed little to no familiarity with technology. This gap was also identified as a concern when discussing cyber-safety in the groups, with participants pointing out the discrepancy in skill levels in courses and computer classes. Many feel apprehensive about hindering the progress of the class due to their limited experience.

\section{Phase Two: Focus Groups}
With the information garnered from the first phase of our study, we implemented a multimodal learning approach to equip older adults with the knowledge they most wanted to learn to feel safer online. Based on the six options given in the dot vote (see Table~\ref{tab:questions-votes}), we selected the three most popular options and created four cyber-safety advice leaflets. 
These advice stemmed from some of the authors' previous research~\cite{murray2023costs,das2020user,das2021organizational,tally2023mid,sheil2022fianan} and established security guidelines, including those from NIST \cite{nist, ncsc}. Each recommendation has been simplified and adapted to be more understandable and practical for individuals with limited technical knowledge. While previous studies, like that by Busse et al. \cite{busse2019replication}, offer security advice for the general public, our aim was to tailor our recommendations to address the specific concerns of older adults. We found that some of the advice from the aforementioned studies was not applicable to this demographic.

These leaflets were presented to two separate focus groups of older adults, totalling 31 participants. Each piece of advice (as mentioned in Table~\ref{tab:cyber-advice}) was accompanied by a short presentation discussing the advice and providing examples which lasted for an average of $20$ minutes. 

\subsection{Analysis}
\label{Analysis}
Two researchers conducted the data analysis and interpretation. One researcher transcribed the focus group audio recordings, centralised the raw data, meticulously reviewed the transcripts, and coded them. The second researcher reviewed the codes to discuss deviations, examine relationships, and uncover prominent themes. After finalising the transcriptions, we analysed user feedback on our tailor-made cyber-safety advice through a three-step process: open coding, axial coding, and thematic analysis. In the open coding phase, we assigned sub-codes to each quote using an inductive coding protocol~\cite{gibbs2007thematic,creswell2017research}. Instead of pre-determined codes, we developed 375 sub-codes from participant quotes, covering emotions, beliefs, attitudes, engagement levels, personal experiences with cyber-crimes, perspectives on conventional cyber-safety advice, digital literacy skills, and feedback on our tailor-made advice. We only sub-coded meaningful and substantial quotes, excluding filler words and short affirmations. We then refined the $375$ sub-codes into $44$ main codes by noting overlaps, grouping similar sub-codes, and retaining unique ones. This axial coding process informed our thematic analysis. After systematically reviewing and coding the dataset with the $44$ main codes, we identified $17$ overarching themes. We also conducted cross-thematic analysis, comparing consistent codes across the four cyber-safety topics to identify overarching themes and patterns.

\subsection{Phase Two: Results \& Discussion }
Overall the response to the advice given was very positive and much like the priority voting exercise, our advice promoted great discussion. 
We will discuss the themes mentioned in Sec.\ref{Analysis} as they relate to the research questions \textbf{\textit{RQ2}} and \textbf{\textit{RQ3}} and the advice given.

\subsubsection{How to avoid scams}
As noted in our dot vote, recognising scams is at the forefront of Irish older adults minds in navigating technology. This was also our most prominent theme (scam and hacking awareness) in our focus groups discussions.
Our participants appear to be aware of mostly all current scams. Despite this, they were keen to see more examples of scams, as they felt that scams are becoming more and more difficult to recognise, 
\begin{quote}
    ``I think all of the examples that you used (in presentation) are relevant, because I've gotten 1 or 2 of them. But I think that scams have become so sophisticated over the last 5 or 6 years that it's very difficult to tell whether something is real or whether it's not, very, very difficult'' (P1). 
\end{quote}
One of our suggestions in avoiding scams was to verify websites via the Check My Link website\footnote{\url{https://check.cyberskills.ie/home}}. This proved to be very popular with many taking note of the website, some older adults were not aware that the padlock symbol in the URL does not completely guarantee the legitimacy of a website. Despite being impressed with the website however,  many felt that they would inadvertently click on the link in order to check; 
\begin{quote}
    ``I love the idea of a website engine check, but you write down the link on paper to do that, and key it in, not engaging with the email or text or whatever'' (P3). 
\end{quote}

While discussing this advice, participants mentioned various scams they had encountered or heard about, including Netflix hacks, Revolut scams, and dating scams. All were very vigilant about answering phone numbers they did not recognise, as P8 noted:
\begin{quote}
    ``I wouldn't be afraid of scams, I think people are far more nervous of answering numbers that you don't know''. 
\end{quote}
Despite being savvy about not answering phone numbers they did not recognise, this has led some to miss important phone calls. Another participant misidentified a notification from his bank as a scam and subsequently his card was blocked.  A common text scam discussed was the fictional son or daughter texting for help and money. Many pointed out that their daughter or son would never ask for money in such a way. A concern for them however was newer technology with AI using their children's voice
\begin{quote}
    ``But they can do the voice now'' (P6).
\end{quote}
A few participants revealed they had their own strategies in dealing with scam texts, for example ensuring that if their children were in real danger they use a safe word only they would know. In this way they can verify if it is indeed them,
\begin{quote}
    ``It's one word out of the title of the program that we were watching. And if the text doesn't in some way contain the word, then it's not from them'' (P1). 
\end{quote} 

\subsubsection{Password management}
Our advice stipulating to keep passwords written down in one safe location was well received. There was a sense of relief about being allowed to write down passwords;
\begin{quote}
    ``That's exactly what I do'' (P2). 
\end{quote}
This also helped with our advice for long passwords, being able to create long and safe passwords, but having concerns about writing them down;
\begin{quote}
    ``What I read was about making strong passwords. But my problem is remembering them, you don't want to write them down'' (P2). 
\end{quote}
Back up for passwords was not common to our participants,
\begin{quote}
    ``Yeah, You're probably safer with your little notebook if you can find your little notebook'' (P2). 
\end{quote}Many of our participants shared passwords with family members and felt comfortable with this. Several individuals reported discomfort with using password managers and having their device remember passwords;
\begin{quote}
    ``And so my concern is exactly that, remembering the password. But, when you save it to your computer and to the cloud, how safe is that? Is it safe at all? To my mind, no, because I don't have control of it'' (P3).
\end{quote}

Many of the participants password habits were unique but their security was questionable;
\begin{quote}
    ``I use a lot of my DOB, and I know it’s very easy to find out'' (P7)\\``I email passwords to myself and keep them there - is that safe?'' (P9)
    \end{quote}
    As well as a false sense of security surrounding password choices;
\begin{quote}
    ``I think I still use the day I got married, so they probably wouldn’t find that out'' (P5). 
\end{quote}
We also found that some commonalities relating to password management were usability and age-related concerns, concerns over password managers online, family assistance and other methods of personal password management. Concerns over password managers has been noted before in previous related work \cite{ray2021older}. One unique method of password management involved a participant using her phone contacts as a password rolladex of sorts,
\begin{quote}
    ``Yea, I put it in like it’s a phone number, so they think it is a mobile number'' (P7). 
\end{quote}When asked what if someone picked up their phone, they responded with-
\begin{quote}
    ``Why would they be looking (at their contacts) ?'' (P7). 
\end{quote}
Usability issues also surfaced around two factor authentication (2FA), relating to anxiety over time limits in inputting the code. They wanted to know more information on this;
\begin{quote}
    ``I’d like some more information on 2FA. It’s something I’ve seen on a number of websites that I’d visit occasionally rather than frequently'' (P2). 
\end{quote}

\subsubsection{Privacy \& cookies}
Privacy was another theme and concern for our participants, noted also by Zou et al. and Zezulak et al.\cite{zou2024cross, zezulak2023sok}. Some of the participants discussed concerns they have;
\begin{quote}
    ``It's getting more difficult, I find. Google is definitely watching [...] they definitely are listening to you. Have you ever found that they're talking about something? And next thing you know, they're on your feed?'' (P8).
\end{quote}
Many appreciated our advice about cookies, as most felt it was necessary to accept cookies at all times,
\begin{quote}
    ``I thought you had to take cookies..my husband said I have to click cookies'' (P5). 
\end{quote}They also expressed curiosity over the purpose of cookies, as well as what to do when they had already accepted cookies;
\begin{quote}
    ``I’ve been on sites where I scroll and it says I have accepted cookies, what do I do then?'' (P9).
\end{quote}
Another participant noted how they liken cookie's purpose to Hansel and Gretel stating;
\begin{quote}
    ``they left bits of cookies along the road to find their way back home; cookies on websites are a way of companies, businesses, websites, etc., finding their way back to you'' (P8).
\end{quote}

Our second piece of advice indicated that some participants were already aware that it is not necessary to fill in every section of forms. One participant mentioned using alternate symbols instead of her phone number. Others found this information very interesting and useful;
\begin{quote}
    ``I didn't realise that you could put x's in your phone number and it would still be accepted, that’s going to make a big difference to me in future'' (P1).
\end{quote}
Another suggestion given to the group was to have back up emails for shopping websites and other subscriptions so they have a main email for important email, 
This generated a lot of interest, and they were eager to learn more about it;
\begin{quote}
    ``[t]hat might be something people may not have thought of doing. A lot of the time people would have things all on the one email account and having a kind of a backup account to use, I wonder is that something you could speak a bit more on'' (P1).
\end{quote}

The last piece of this sections advice addressed avoiding sharing personal information online. There was a general awareness of this fact also. For example, one participant voiced her shock at how they had observed on RIP.ie~\footnote{\url{https://rip.ie/} is an Irish website which lists recent deaths and notices regarding funeral details.} that a neighbour had disclosed a persons address in the comments;
\begin{quote}
    ``I read the condolences, a neighbour of mine had died. And somebody put in the condolences. And they gave her a number of her house and her address unknowingly'' (P6).
\end{quote} 

\subsubsection{What to Do if You are Hacked}
Our last piece of advice concerned solutions to being hacked. This was an area that our participants were most curious about. They felt there was not much options or help if and when they were hacked or scammed. On the subject of reporting a crime and the PULSE ID\footnote{PULSE number is a computer-generated identifier for a crime or incident, used by the Gardaí to track the case.}, there were conflicting opinions.

One participant (P4) recalled an occasion where they had to report an online hacking and noted their Garda station had a specialised person who helped with people who are not proficient online and found her extremely helpful. However this was just one case, most appeared sceptical of the attention of the Gardaí to their plight in such cases
\begin{quote}
    ``[I] would wonder whether the guards behind the desk would think it was important. Would they automatically give you a PULSE ID, not think twice about it?'' (P1). 
\end{quote}
Despite this, they showed interest in the PULSE ID and wanted to know more;
\begin{quote}
    ``I think that’s a big thing to know about'' (P1).
\end{quote}
P3 further elaborated;
\begin{quote}
    ``This is wonderful. This should be replicated in every community centre in age action groups or whatever country, because people just aren't aware of the rights they have''. 
\end{quote}
Our advice on hacking appeared to be of great use to our participants,
\begin{quote}
    ``The information has been very sharp, very concentrated, very spot on. It's been wonderful''(P3).
\end{quote}
Banking online is a big fear for many, as they feel that going online with your finances is too risky and exposed. 
There was confusion regarding when a transaction was completed and whether their funds were at risk of being compromised during online transactions. Many mentioned Revolut as a major concern
\begin{quote}
    ``I’m not convinced of Revolut now, I know a lot of people love it, it’s getting a bit different now'' (P6).
\end{quote}

\subsection{Limitations and Future Work}
Our study revealed valuable insights into the cyber-safety needs of older adults, but it is not without limitations. The focus group setting may have inhibited some participants from expressing their thoughts publicly, and managing overlapping discourse made accurate transcription challenging. To address these limitations, future research could administer follow-up questionnaires using a 10-point Likert scale to rate the usefulness of the advice and gather additional suggestions. We plan to conduct further studies where we will also consider tailoring materials to different tech levels among older adults and exploring each cyber-safety topic in greater detail. Additionally, expanding the scope to include diverse demographic groups could provide a more comprehensive understanding of the digital divide and help develop more inclusive educational materials.

\section{Conclusion}
The digital divide continues to affect older adults, limiting their ability to navigate and benefit from the online world. Our two-phase study aimed to address this issue by identifying the cyber-safety priorities of older adults in Ireland and developing tailored educational materials to enhance their online safety. In Phase I, we used Dot Voting with $58$ participants to determine their top cyber-security concerns, which included password management, privacy, and avoiding scams. These priorities informed the creation of four pieces of cyber-safety advice that were then presented in Phase II to $31$ participants through focus groups. The response to the advice was overwhelmingly positive. Participants valued practical tools for verifying website legitimacy and email breaches, the use of passphrases, and guidance on cookies and PULSE IDs for reporting scams. The focus group discussions highlighted a strong desire among older adults to stay informed about evolving scams, managing hacking attempts, using password managers, understanding 2FA, and banking safely online. There was also significant interest in learning more about cookies and managing online privacy.
Our study demonstrates that older adults can greatly benefit from tailored cyber-safety education, which enhances their confidence and competence in navigating the digital landscape. 

\begin{acks}
    This publication has emanated from research conducted with the financial support of the EU Commission Recovery and Resilience Facility under the Science Foundation Ireland OurTech Challenge Grant Number 22/NCF/OT/11212. Dr Murray is supported in part by a research grant from Science Foundation Ireland (SFI) and is co-funded under the European Regional Development Fund under Grant Number 13/RC/2077\_P2. This research was supported by a Google Trust and Safety Research Award.
\end{acks}

\bibliographystyle{ACM-Reference-Format}
\bibliography{bibliography}

\end{document}